\documentclass[lettersize,journal]{IEEEtran}
\usepackage{amsmath,amsfonts}
\usepackage{algorithmic}
\usepackage{algorithm}
\usepackage{array}
\usepackage[caption=false,font=normalsize,labelfont=sf,textfont=sf]{subfig}
\usepackage{textcomp}
\usepackage{stfloats}
\usepackage{url}
\usepackage{verbatim}
\usepackage{graphicx}
\usepackage{cite}
\usepackage{ulem}
\begin{document}

\title{Antenna Pattern Modelling Accuracy for a Very Large Aperture Array Radio Telescope with Strongly Coupled Elements
}

\author{Pietro Bolli, 
\thanks{INAF-Osservatorio Astrofisico di Arcetri, Florence, Italy} 
David Davidson,~\IEEEmembership{Fellow,~IEEE},         %
\thanks{International Centre for Radio Astronomy Research, Curtin University, Perth, Australia} %
Maria Grazia Labate,    %
\thanks{Square Kilometre Array Observatory, Manchester, UK} %
Stefan J. Wijnholds,~\IEEEmembership{Senior Member,~IEEE}         %
\thanks{Netherlands Institute for Radio Astronomy (ASTRON), Dwingeloo, The Netherlands} %
\thanks{As we consider the contribution of each author instrumental to this work, we have listed our names in alphabetical order.}
\thanks{Manuscript received TBC ; revised TBC.}}

\markboth{IEEE Antennas and Wireless Propagation Letters,~Vol.~X, No.~Y, Month~YYYY}%
{Shell \MakeLowercase{\textit{et al.}}: A Sample Article Using IEEEtran.cls for IEEE Journals}



\maketitle

\begin{abstract}Modern radio telescopes strongly rely on accurate computational electromagnetic tools for ``beam'' models. Especially for densely-packed aperture array radio telescopes, the only feasible way to produce accurate models of the individual embedded element patterns is by using electromagnetic codes. In this paper, the accuracy of two models computed by different commercial codes is evaluated for one station of the SKA-Low radio telescope. Except for a couple of critical frequencies, the amplitude and phase errors are low enough to allow a beamformer efficiency higher than 99\%.
\end{abstract}


\section{Introduction}

\begin{figure}
    \centering
    \includegraphics[width=\columnwidth]{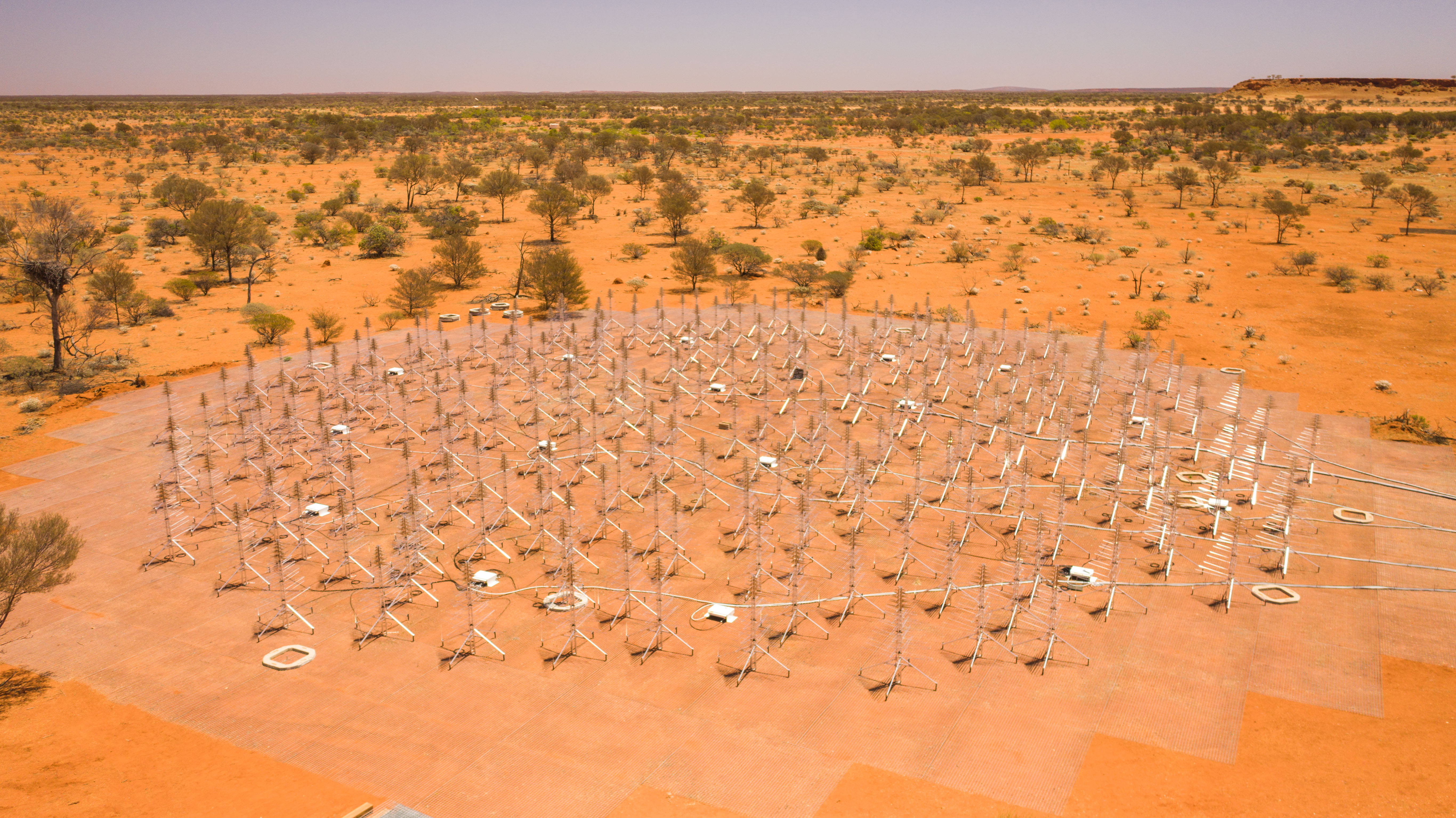}
    \caption{An aerial view of the Aperture Array Verification System 2 (AAVS2), an SKA-Low prototype station deployed in 2020 at the Inyarrimanha Ilgari Bundara, the Murchison Radio-astronomy Observatory site. Credits: ICRAR/INAF.}
    \label{fig:AAVS2}
\end{figure}

\IEEEPARstart{T}{he} Square Kilometre Array (SKA) Observatory is currently engaged in the construction of two telescopes. SKA-Mid is a dish-based system, operating from 350~MHz to around 15~GHz \cite{Swart2022JATIS}, and SKA-Low is an aperture array-based system, operating from 50 to 350~MHz \cite{Labate2022JATIS}. SKA-Low will comprise several hundred ``stations'' (subarrays) of approximately 40~m diameter; each station is an array of 256 dual-polarized log-periodic SKALA4.1 antennas \cite{Bolli_etal_2020}. An aerial photograph of the current SKA-Low prototype station is shown in Fig.~\ref{fig:AAVS2}. Especially at the lower frequencies, the antennas are electromagnetically in close proximity, and mutual coupling effects are significant \cite{BolliDavidsonJATIS2022}.

The new generation of low-frequency radio telescopes currently under construction are likely to rely on simulated patterns to a far greater extent than their predecessors.  Traditionally, radio telescopes have been calibrated using measured beams. However, instruments such as SKA-Low are characterised by an ``all sky’’ response, very different to the narrow beams of dishes, and there is a scarcity of sufficiently bright astronomical calibrators  to individually calibrate the low-gain antennas comprising the station.

Initial commissioning results \cite{Wijnholds2020GASS,Wayth_PAFAR2022} for the SKA prototype array considered in this paper indicate that using individual Embedded Element Patterns (EEPs) will be needed to obtain the good station calibration solutions towards the lower end of the operating band. Furthermore, accurate prediction of the station beam using the EEPs plays a crucial role for the overall performance of the telescope and in particular for the beamformer efficiency. All these considerations motivate the importance of evaluating simulated EEP accuracy.

Modelling a large and electromagnetically complex real-world system, such as the SKA-Low station shown in Fig.~\ref{fig:AAVS2}, requires a number of engineering trade-offs. From a simulation viewpoint, this is a very challenging problem. Initial work, around a decade back, on the full-wave modelling of early SKA-Low station prototypes, involved a much smaller number of simpler and physically smaller antennas \cite{Sutinjo_etal2015}. This early work has developed into the current workflow \cite{BolliDavidsonJATIS2022}; simultaneous advances in code capability and High Performance Computing (HPC)  systems have made the present ``digital twin'' just realizable at the time of writing. The SKALA4.1 antenna element alone requires around 10~000 degrees of freedom to adequately model with Method of Moments (MoM) codes, while modelling all 256 antennas requires well in excess of a million unknowns. As the antenna is primarily a wire structure, the MoM, accelerated by the Multi-Level Fast Multipole Method (MLFMM) to permit modelling the entire station, is the obvious choice of simulation technique.

The desired accuracy required in the modelling of antenna patterns is not a straightforward decision.As outlined above, a model satisfying the usual meshing guidelines pushes the limits of current modelling tools and HPC systems in terms of run-time and memory, and also in terms of post-processing. Moreover,  other factors impacting on the overall accuracy have to be taken into account within the  operational context. For example, distortions introduced by the ionosphere may prove more significant than limitations in pattern accuracy.

In the context of the SKA-Low telescope, the accuracy really needed should be sufficient to firstly predict critical key performance metrics, such as sensitivity and polarization purity, in an environment as close as possible to the real one; and secondly, to enable calibration at both station and telescope array level to be performed \cite{Labate_etal2023}. 

The present paper extends comparisons in \cite{BolliDavidsonJATIS2022} between two different Computational Electromagnetics (CEM) models and codes to include phase, which is clearly crucial for beamforming. In addition to direct comparisons of amplitude and phase computed by two different models, beamforming efficiency is introduced as a relevant metric for assessing model accuracy for  radio telescopes comprising aperture arrays.

 We present the differences in terms of both direct comparison of the computed fields, as well as via the errors in a beam formed by a conjugate field-matched polarimetric beamformer. The latter allows us to assess the errors for signals with polarization matched to one of the feeds as well as unpolarized signals. We conclude that the numerical errors are sufficiently small at most frequencies in the SKA-Low operating frequency range to have a small or even negligible impact on beamformer efficiency.

\section{Tolerance on beamforming errors}

The SKA-Low station uses a digital beamformer \cite{Labate2022JATIS}. Beamformer efficiency is defined in \cite{Comoretto_etal2020}. Phase and amplitude errors in the station beamformer result in a reduction in beamformer efficiency that incurs a sensitivity loss. If the beamformer efficiency of the stations is 99\%, the overall sensitivity of the telescope is reduced by 1\% compared to an ideal array of stations with a beamformer efficiency of 100\%. In the case of SKA, a 1\% loss in beamformer efficiency is therefore equivalent to losing 5 (of the 512) stations. This argument enables an engineering trade-off between the cost of improving the beamformer efficiency of the individual stations and the cost of building more stations to achieve the desired telescope sensitivity at minimal cost.

As the main beam directivity of the stations is proportional to the beamformer efficiency, the average side lobe level (SLL) of a pseudo-random array, like the AAVS2 station, increases proportional to the decrease in beamformer efficiency. The beamformer efficiency thus provides a good measure of their impact on overall array performance for SKA-Low. An overview of the beamformer error budget is presented in \cite{Wijnholds2022ATAPRASC}. In this work, we concentrate on just one of the terms in the error budget, being the accuracy of the EEP models, so the maximum tolerable error is a loose upper bound on that error.

Generic expressions for the impact of beamformer errors are presented in various works \cite{Mailloux2022, Wijnholds2020GASS}. As there are different underlying factors for amplitude and phase errors, it is useful to consider them separately, thus providing a budget to each of them. Based on \cite[Eq.~(7.13)]{Mailloux2022}, the impact of amplitude and phase errors with standard deviation $\sigma_\mathrm{A}$ and $\sigma_\phi$ respectively on beamformer efficiency can be described by
\begin{equation}
\eta_\mathrm{BF} = \frac{1}{1 + \sigma_\mathrm{A}^2 + \sigma_\phi^2} \approx 1 - \sigma_\mathrm{A}^2 - \sigma_\phi^2,
\label{eq:BF_efficiency}
\end{equation}
where the approximation is based on the first terms of the Taylor expansion assuming that the errors are small. It is also assumed that the nominal amplitudes are normalized to unity and that phases are expressed in radians. If the error budget is evenly distributed between amplitude and phase errors, this result is consistent with \cite[Eq.~(5)]{Wijnholds2020GASS}.

\section{Evaluating errors in numerical simulations}

By their nature, numerical simulations contain errors. The most common sources of error include: unconverged solutions (the mesh is not sufficiently fine); inaccurate geometrical models (the computational model differs from the as-built antenna, especially in the excitation ports); limitations of the tools (e.g., modelling of finite meshed ground plane above a real ground as an infinite Perfect Electric Conductor (PEC) ground or assuming the antenna material as a PEC); and slow  convergence of the iterative solvers some fast methods rely on. 

\begin{figure}
    \centering
    \includegraphics[width=0.4\columnwidth]{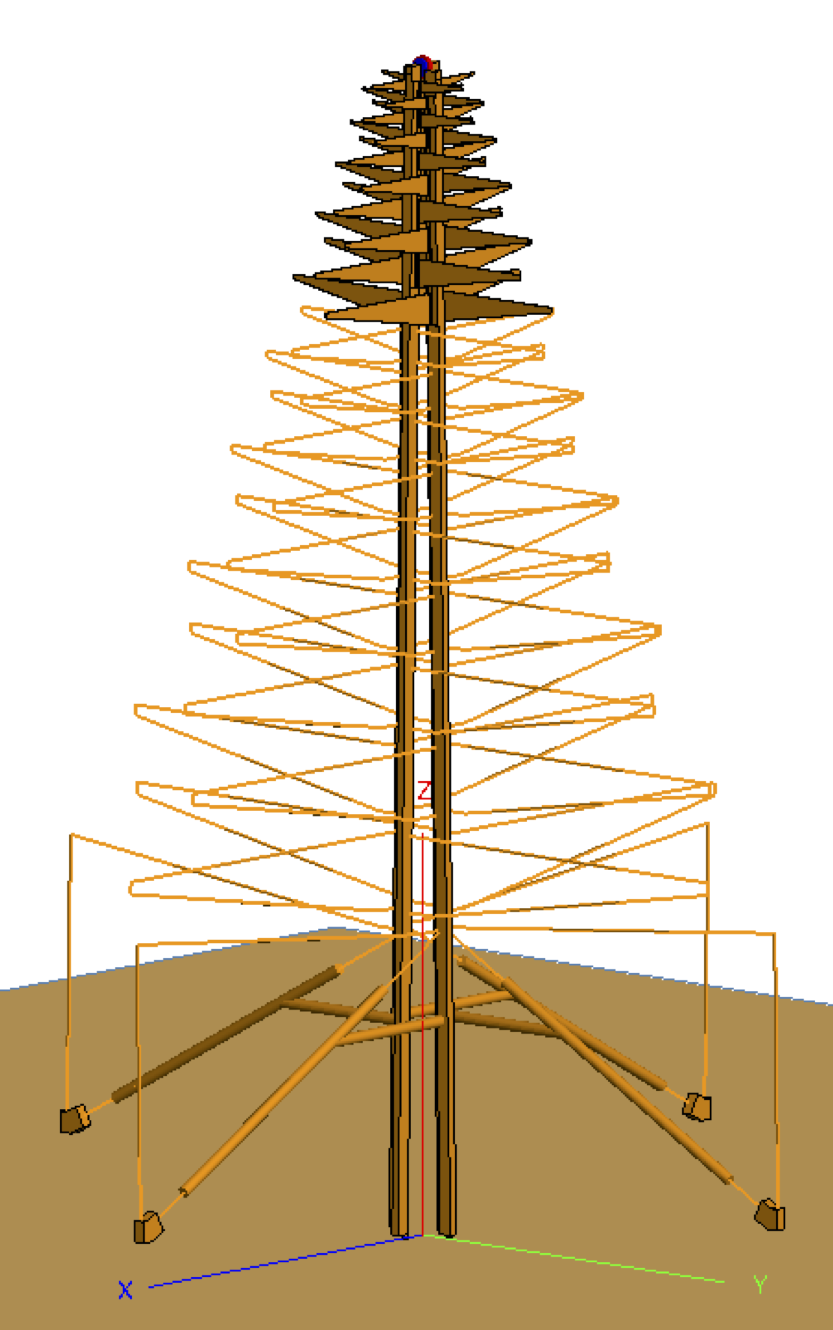}
        \includegraphics[width=0.4\columnwidth]{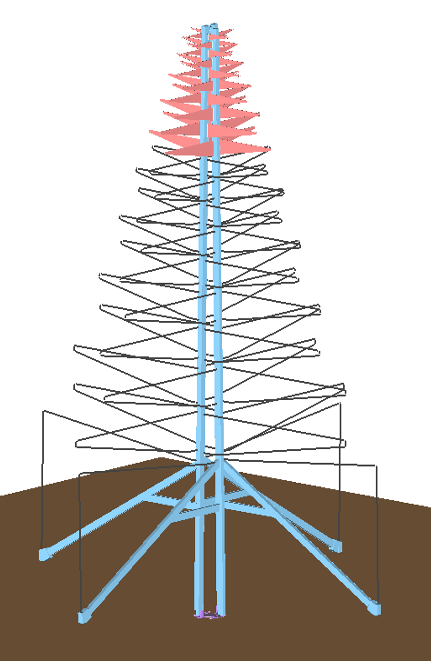}
    \caption{The CAD models for FEKO (left) and Galileo (right). Whilst visually these appear almost identical, there are subtle differences in meshing and feed which result in differing solutions.}
    \label{fig:CADmodels}
\end{figure}

The rigorous evaluation of errors in computational electromagnetic solutions is a challenging mathematical topic. This is especially so for integral-equation based full-wave methods, such as the MoM used to compute the EEPs. Firstly, few problems admit analytical solutions. Secondly, the solutions can require mathematical constructs  unfamiliar to most RF engineers \cite{MarchandDavidson2014,BuffaChristiansen2003}. 


A pragmatic approach to establishing errors is to check convergence, but the aperture arrays under consideration here are electrically so large that it is not possible to refine the mesh beyond an adequate initial mesh. Convergence checks were indeed performed on individual elements, see \cite{DavidsonBolli_etal_ICEAA19}. Comparison against measured data is also very difficult, as these arrays are too large to be deployed in an antenna chamber. Measurements using a probe mounted on an Unmanned Aerial Vehicle (UAV) were undertaken during construction of the AAVS2 station \cite{Virone2020}, but the experimental procedure is subject to inaccuracies as well, such as position and instrument errors, and is very costly. The accuracy obtainable by such systems is further compromised by the remote and hostile environment of the site. The aforementioned campaign, for example, was impacted by high winds. 

The approach taken here is to use two different electromagnetic simulators - FEKO and Galileo. Both are MoM codes, with the MLFMM accelerator; see, for instance, \cite{Davidson2} for a description of these methods. Examples of the CAD models simplified for FEKO and Galileo simulations are shown in Fig.~\ref{fig:CADmodels}. Some of the assumptions made in the CEM models are discussed in \cite{BolliDavidsonJATIS2022}. 

\section{Results}

In previous work \cite{BolliDavidsonJATIS2022}, results for SKALA4.1 gains and overall station beams have been presented for AAVS2. In that work, the focus was largely on the amplitude errors of the EEPs. However, for beamforming, phase errors will be also a significant source of error. In  Table~\ref{tab:table1}, we compare magnitude (amplitude) and phase errors between the FEKO and Galileo solutions for the 256 dual-polarized antennas of AAVS2 (as in \cite{BolliDavidsonJATIS2022}) at zenith and along the principle planes (H- and E-plane, respectively) from $\theta=-45^\circ$ to $\theta=45^\circ$ (where $\theta$ is the angle from zenith). Frequencies have been selected to  sample the working band of SKA-Low and also capture some critical frequencies.  For magnitude errors, some form of normalization is required, and the error at each angle is normalized by the average of the FEKO and Galileo EEP maximum amplitude at zenith. For phase errors, the result is  the standard deviation of the difference between the phase of the FEKO and Galileo voltage gains at zenith, or in the relevant cut plane. Weighting the phase errors in some fashion --- e.g.\ using the ratio between local and maximum amplitude --- was considered but not used, as phase errors impact directly on the formed beam. 
In Table~\ref{tab:table1}, the results are for the co-polarized field component, and these are for the ``Y'' polarization, (feed 1) --- dipole arms parallel to the north-south direction. 

\begin{figure}
    \centering
    \includegraphics[width=\columnwidth]{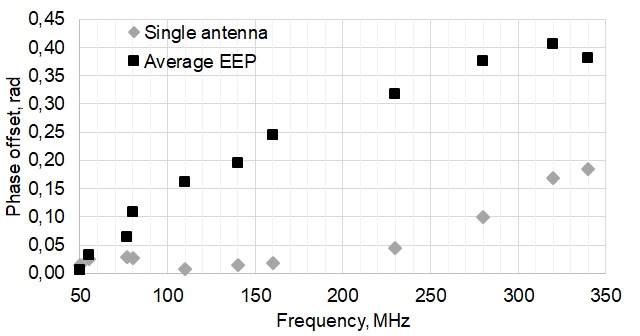}
    \caption{Phase difference between FEKO and Galileo at zenith for the single antenna model and for the mean EEPs.}
    \label{fig:ph_er}
\end{figure}

Besides this error variability, we have previously noted a constant offset along any direction in the sky in EEP phase solutions between these codes. This phase error is shown in Fig.~\ref{fig:ph_er} at zenith for the isolated SKALA4.1 antenna and for the average of EEPs comprising AAVS2. It is evident that in both cases the error increases with frequency and also that for the EEPs the error is much larger than for the single antenna. The reason for this offset is most likely in how the feed region is meshed and modelled by the two solvers, and this brings a degree of variability in a very sensitive region. This is even further emphasized for the multi-antenna configuration, where the mutual coupling increases the complexity of the EM problem. Fortunately, this offset is readily removed during calibration and therefore is neglected here after.

What is notable from Table~\ref{tab:table1}  is the overall good agreement of both the amplitude and phases in both principal planes, with two exceptions. This agreement is  generally within tens of milliradians (several degrees) at most for phase, or a few percent for amplitude, other than at a two specific frequencies (55 and 77~MHz). At these frequencies, errors can be an order of magnitude larger. These frequencies are known ``problem'' frequencies for simulation, due to mutual coupling being especially pronounced \cite{Bolli2022}.  At the time of writing, EEPs have been computed at 5 MHz cadence over the entire operating band, and at 1 MHz intervals from 70--90 MHz. Only a few of these problematic frequencies were noted and this is accompanied by convergence issues with the MoM MLFMM solvers, which flags potential accuracy issues. Finally, E-plane results are in general worse than in the H-plane, which is likely due to a null in the field pattern within 45 degrees off zenith. (For instance, the E-plane result at 55~MHz is not reported as the agreement is too poor to permit a useful comparison).

There is a well-known rule of thumb in the phased array community \cite[Eq.~7.10]{Mailloux2022} which gives the equivalent phase error expressed in degrees (for an overall pattern error in terms of residual sidelobes) as 6.6 times the amplitude error in dB, and the results in these figures are broadly in this regime, indicating a balance between amplitude and phase errors. 


\begin{table}[!t]
\caption{Standard deviation of  errors between FEKO and Galileo solutions for all 256 antennas. Magnitude is normalised linear; phase is in milliradians. \label{tab:table1}}
\centering
\begin{tabular}{|c||c|c|c|c|c|c|}
\hline
Freq. & \multicolumn{6}{c|}{Standard deviation}   \\
(MHz)     & \multicolumn{2}{c|}{zenith}  & \multicolumn{2}{c|}{H-plane} & \multicolumn{2}{c|}{E-plane}\\ 
 & mag. & phase & mag. & phase & mag & phase \\ 
 &      & (mrad)&      & (mrad)&     & (mrad)\\\hline
50  & 0.008  & 6.46 & 0.007 &  6.28 & 0.008 & 38.4 \\
55  & 0.213  & 187  & 0.175 &  211  & 0.239 & ---  \\
77  & 0.136  & 175  & 0.138 &  197  & 0.117 & 1380 \\
80  & 0.008  & 9.25 & 0.009 & 11.7  & 0.046 & 119  \\
110 & 0.010  & 9.25 & 0.010 & 12.0  & 0.014 & 20.6 \\
140 & 0.026  & 25.0 & 0.028 & 33.9  & 0.040 & 63.0 \\
160 & 0.017  & 16.1 & 0.016 & 19.7  & 0.023 & 35.4 \\
230 & 0.038  & 40.0 & 0.034 & 40.5  & 0.034 & 50.8 \\
280 & 0.018  & 20.4 & 0.017 & 21.8  & 0.022 & 32.3 \\
320 & 0.023  & 24.3 & 0.024 & 29.7  & 0.028 & 40.3 \\
340 & 0.017  & 21.1 & 0.020 & 26.9  & 0.025 & 37.2 \\
\hline
\end{tabular}
\end{table}

\begin{figure}
    \centering
    \includegraphics[width=\columnwidth]{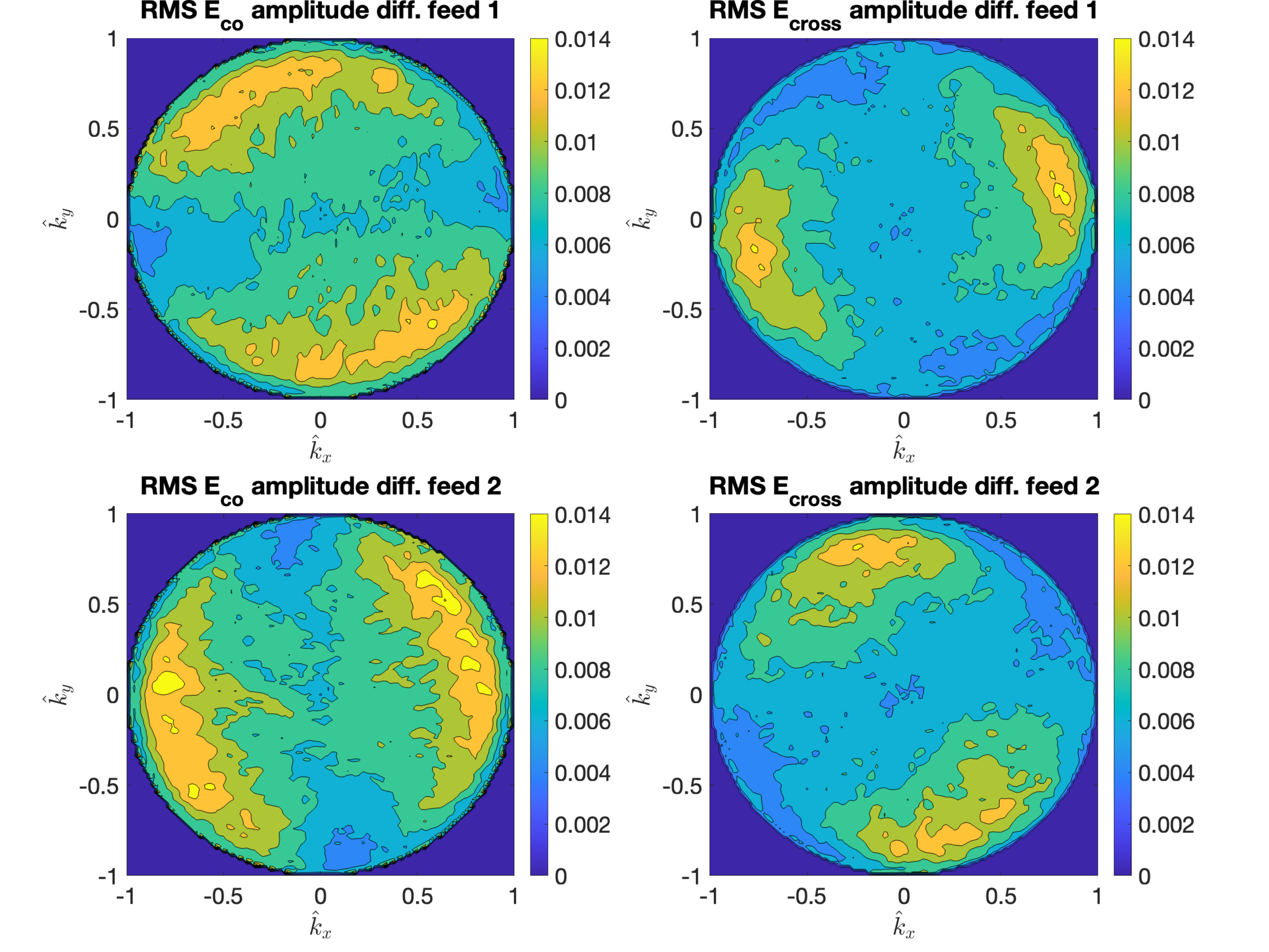}
    \caption{RMS amplitude difference  at 110 MHz between the models of the 256 individual EEPs across the sky for both antenna feeds and for the indicated polarizations.
}
    \label{fig:rms_diff_ampl}
\end{figure}

\begin{figure}
    \centering
    \includegraphics[width=\columnwidth]{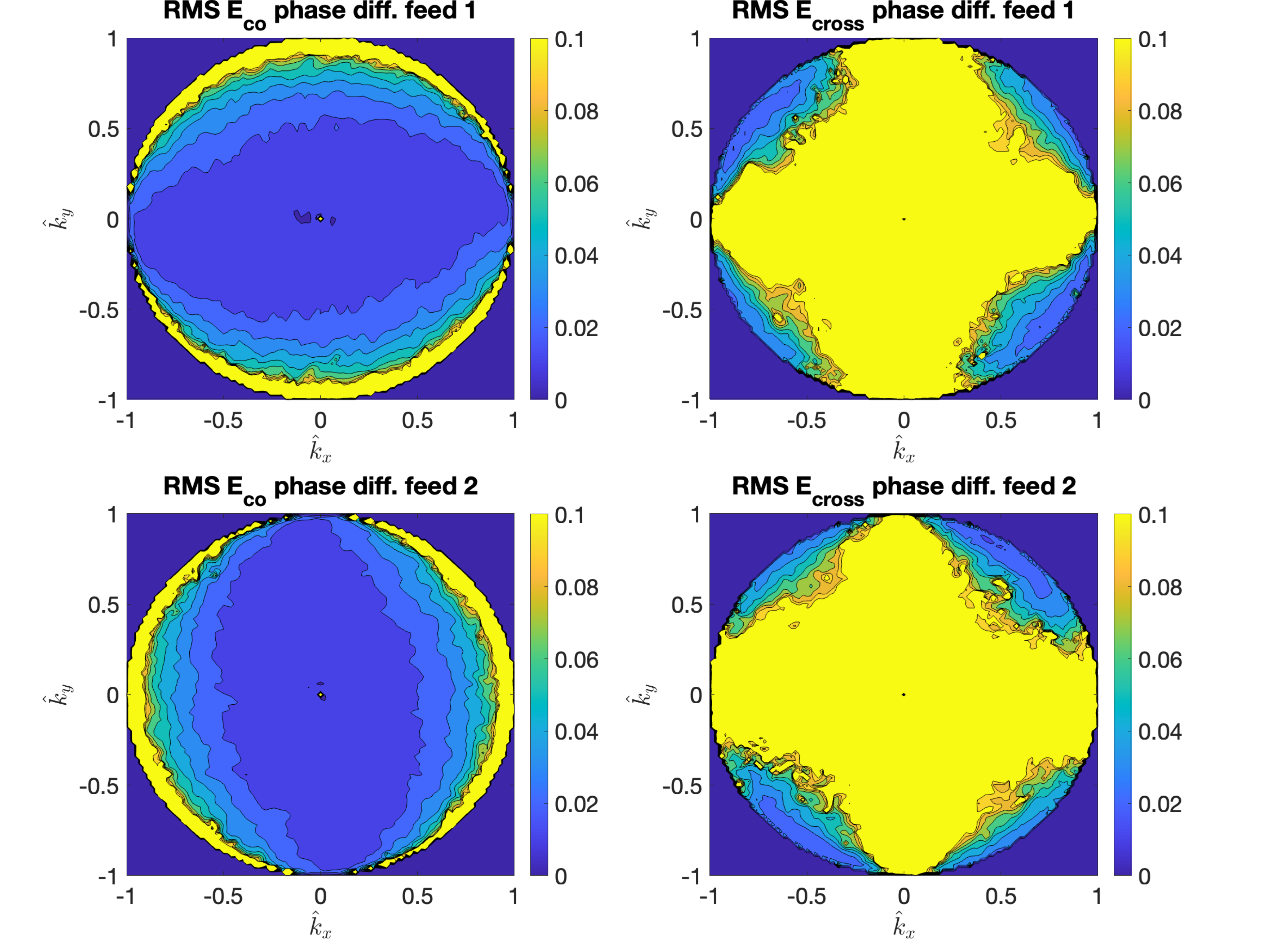}
    \caption{RMS phase difference (in radians) at 110 MHz between the models of the 256 individual EEPs across the sky for both antenna feeds and for the indicated polarizations.}
    \label{fig:rms_diff_phase}
\end{figure}

Above, we discussed the modelling errors towards boresight and in the H-plane and E-plane. To assess the error across the sky, we have calculated the RMS difference between EEPs calculated by Galileo and FEKO for a grid of directions at 110 MHz. We chose this frequency as it is representative of the two operational modes of the arrays, viz.\ dense and sparse regimes --- it is on the boundary between them.  The RMS differences in amplitude and phase are presented in Figs.~\ref{fig:rms_diff_ampl} and \ref{fig:rms_diff_phase} respectively. Results are presented using the Ludwig-II polarisation convention for the two orthogonal feeds \cite{Ludwig1973}.

The amplitude errors shown in Fig.~\ref{fig:rms_diff_ampl} are expressed on a linear scale as a fraction of the maximum gain (normalized to unity). These errors are directly comparable to the fourth and sixth columns in Table~\ref{tab:table1} for 110~MHz. The former should be compared to a vertical cut at $\hat{k}_x=0$ on the upper left-hand sub-plot in Fig.~\ref{fig:rms_diff_ampl}, and the latter to a horizontal cut at $\hat{k}_y=0$ on the same sub-plot, noting also that the table only includes errors out to $\pm 45^\circ$. Due to the normalization, the amplitude errors are lowest in the directions in which the antenna gain is lowest for the polarization concerned. 

The phase errors shown in Fig.~\ref{fig:rms_diff_phase} are expressed in radians. For readability, the highest contour level was defined at 0.1 radians ($\sim 6^\circ$). In the directions where the gain is low, especially for the cross-polarization component, small differences can lead to very large phase errors, which explains the overall structure of these plots. Fortunately, those directions will get a very low beamformer weight in a field-matched polarimetric beamformer. A more detailed look at the dark blue areas indicates that the RMS error near boresight is in the order of 0.01 radians, which corroborates our earlier finding that there is a balance between the amplitude and phase errors.

\begin{figure}
\centering
\includegraphics[width=0.49\columnwidth]{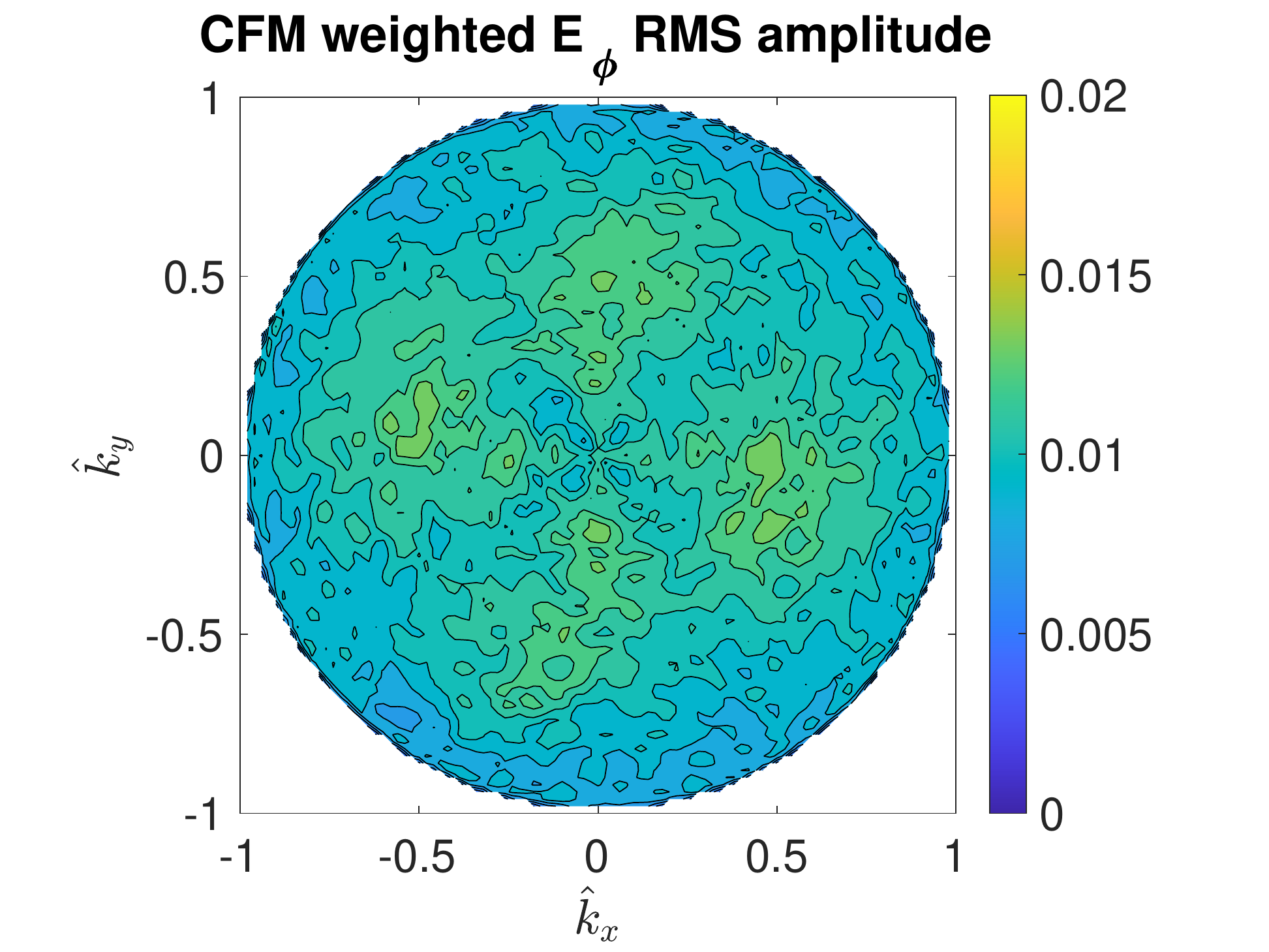}
\includegraphics[width=0.49\columnwidth]{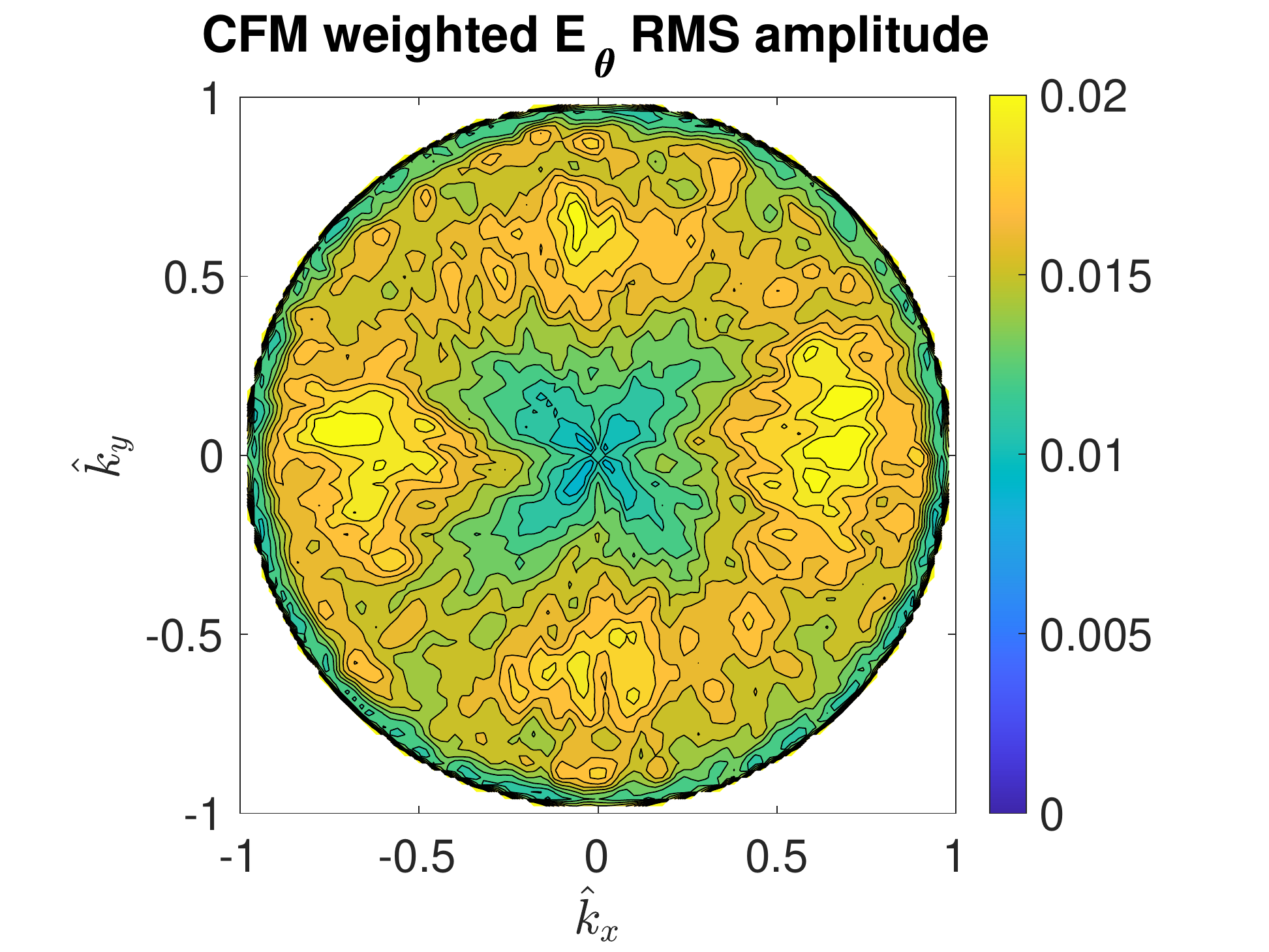}
\caption{CFM weighted RMS amplitude errors across all 512 feeds for $E_\phi$ and $E_\theta$ polarization respectively at 110MHz. \label{fig:CFM_weighted_RMS}}
\end{figure}

In such a conjugate field-matched (CFM) polarimetric beamformer  \cite{Veen2019,Warnicketal}, the highest beamformer weights will be assigned to the feeds that have the highest gain for the direction and polarization concerned, so the most relevant amplitude errors are those within the desired scan range ($\pm 45^\circ$ from zenith) along the direction with highest gain. This is confirmed by the CFM weighted RMS amplitude errors for the $E_\phi$ and $E_\theta$ response of the array across all 512 feeds shown in Fig.~\ref{fig:CFM_weighted_RMS}.

\section{Conclusions}

We assessed the accuracy of the EEP models for the SKA-Low stations by comparing the results of two state-of-the-art commercial CEM simulation tools, Galileo and FEKO. We found that the RMS difference between these CEM simulators is less than a few percent in magnitude and $\sim 0.05$ radians ($3^\circ$) in phase over the area where the elements have significant gain (for all except two known ``problem'' frequencies). With that level of errors, a beamformer efficiency higher than 99\% is achievable. Given other practical considerations, such as production tolerances and noise in calibration observations, we thus feel confident that we have cross-validated our CEM simulations to a suitable level for beamforming applications.

Continuing advances in this field may yield even more capable solution techniques  in future years \cite{Gueuning_etal2022,Conradie2022}, permitting greater modelling fidelity should these methods become available in commercial codes. 

\section{Acknowledgments}

The EEP datasets used in this work were computed by D. Ung (ICRAR-Curtin, Australia) and M. Bercigli (IDS, Italy). Some of this data was computed on the Pawsey supercomputer centre and on DUG's commercial HPC systems, both located in Perth, Western Australia. This work was supported by the Foundation for Dutch Scientific Research Institutes.

\bibliographystyle{./bibliography/IEEEtran}
\bibliography{./bibliography/IEEEabrv,./bibliography/bibliog}

\end{document}